\documentstyle[proceedings]{crckapb}                   
                                        
\begin{opening}                         
\title{ACCELERATION OF ULTRA HIGH ENERGY COSMIC RAYS:\protect\\            
       COSMIC ZEVATRONS?}        
\author{T.W. JONES}                
\institute{Department of Astronomy\\                
           University of Minnesota, Minneapolis, MN, 55455 USA}           
                                        
\end{opening}                           
                                        
\runningtitle{Acceleration of UHECRs}   
                                        
\begin{document}                        
\def\lsim{\raise0.3ex\hbox{$<$}\kern-0.75em{\lower0.65ex\hbox{$\sim$}}}
\def\gsim{\raise0.3ex\hbox{$>$}\kern-0.75em{\lower0.65ex\hbox{$\sim$}}}
\def\etal{{\it et al.~}}
                                        
\begin{abstract}

In this lecture I outline some of the underlying physics issues associated with 
accelerators plausibly capable of explaining the UHECRs up to ZeV energies. I concentrate on the 
mechanisms and their constraints, but provide a brief background on
observations and the constraints they supply, as well. 
\end{abstract}
                                        
\section{Introduction}                 

Strong observational evidence now exists for the
existence of cosmic rays (CRs) at energies approaching 1 ZeV (=$10^{21}$eV).
Their origins remain a mystery and a serious theoretical challenge.
In this lecture I will outline some of the key issues
that constrain theories for production of these ``Ultra High Energy
Cosmic Rays'' (UHECRs). I will focus on so-called ``bottom-up'' scenarios,
which accelerate charged particles, usually protons, from low energies.
These hypothetical accelerators have been dubbed ``Zevatrons'',
since they must achieve ZeV energies.
Other lecturers at this School have addressed the alternative
``top-down'' scenarios, which explain UHECRs as decay products
of fossil Grand Unification defects, or by 
``new'' physics (e.g., Stecker \& S\'anchez). 
The new physics approaches also include explanations
that reduce constraints on long range UHECR propagation by allowing violation of
Lorentz invariance, so those approaches can combine both scenarios. 
In practice, both bottom-up and top-down scenarios are generally forced to fairly
extreme model assumptions in order to account for current measurements. So, which
ever approach turns out to be most relevant, we will learn some very interesting
physics or astrophysics. The coming generations of cosmic ray detectors
should tell us which direction to go within the next few years.

In addition to the other lectures from this School readers interested
in more details or alternative perspectives may want to read
some of the excellent reviews of this topic that have appeared recently in
the literature (e.g., Olinto 2000; Watson, 2000; Sigl, 2002).

I begin my discussion with a short outline of some background that
helps to define the underlying issues. My main goal is to confront
the physics that must be included in cosmic Zevatrons, so
\S 3 aims to find the underlying constraints that can be used to
limit the field of models.
Since many models accelerate particles in shock
waves, I will follow with a short discussion of the physics of
diffusive shock acceleration, before turning to apply the
appropriate physical constraints to some candidate phenomena in \S 5.

\section{Some Background}     

The cosmic ray energy spectrum has been measured over almost
13 decades in energy, roughly between $\sim 0.1$GeV and $10^{12}~$GeV
( = 1 ZeV).
At lower energies the solar wind traps extra-solar-system charged particles and
carries them away. At higher energies the extrapolated fluxes are extremely
small at best ($< 10$ particles km$^{-2}$ century$^{-1}$), so not yet 
measured or adequately constrained.
To a first approximation the measured energy spectrum is a broken powerlaw with a slope
close to -2.7 below a ``knee'' near $10^6$GeV, then steepening by
about 0.5 in the index above the knee. The spectrum appears to flatten slightly
again above $\sim 10^9$GeV (= 1 EeV). This feature is sometimes called the
``ankle'' of the CR spectrum. The knee and ankle features are commonly
viewed as corresponding to transitions either in the nature of
the CR accelerator or in the propagation of the CRs. 
As outlined below there is currently a controversy about
the spectrum above 100 EeV (Abu-Zayyad, \etal 2002; Takeda, \etal 2002), but several groups have independently
detected events with energies in this range, so there is little
doubt about their existence. The highest energy event reported so
far is 0.3 ZeV (Bird, \etal 1994). 
That amounts to an astounding 50 Joules of kinetic energy in an individual
elementary particle.
Not only is it problematic to explain the generation of CRs at
these high energies, their propagation is severely limited, especially by
interactions with the cosmic microwave background (CMB). The difficulties
and implications are obviously very important, so there are
major efforts underway to improve on current measurements.

The composition of the 
cosmic rays near or above the knee is not well established. As
discussed by Stecker in his contribution at this School (Stecker 2002)
the evidence mostly supports a transition from a
mix of light and heavy nuclei near and just above the knee to mostly light
nuclei at the highest energies. Most, but not all discussions of UHECRs
take them to be protons for that reason. Additionally, collisions
between ultrahigh energy heavy ions and photons disrupt heavy ions,
so that they cannot propagate over long range. (See Stecker 2002 for
a discussion of these properties.)

Another important property of the UHECRs that constrains their origins 
or possibly their propagation is
the distribution of detected events on the sky.  Below the knee CR
detections
are nearly isotropic (e.g., Erlykin, \etal 2002), but that is a reflection
of that fact that low energy CR propagation is diffusive in the turbulent
galactic magnetic field.  Sources of these CRs are obscured because of that.
On the other hand, since the gyroradius of a particle with charge $Z$
and energy $E_{EeV} = E/$EeV is 
$r_g \approx (E_{EeV}/ZB_{\mu G}~$kpc),
and the characteristic galactic magnetic field is a few $\mu$G,
confinement of CRs within the galaxy is limited to energies below
$\sim 1 E_{EeV}/Z$. Diffusive propagation within the galaxy
should break down well below
that energy. Indeed, around the ankle some enhancement in the
CR flux is reported towards
the galactic center (Hayashida, \etal 1999), so we may have
seen direct indications of their galactic origins. At higher energies, however,
the data are consistent with isotropic detections (Takeda, \etal 1999;
Uchihori, \etal 2000). Unless the 
UHECRs are highly charged ($Z>>1$) this almost assures their extragalactic origins.
                                        
\subsection{Some Propagation Issues}                  

Before discussing the physics of various acceleration scenarios
it is helpful to establish a clear picture of what general limits
can be placed on the locations of possible extragalactic sources of the
UHECRs.  The most frequently discussed constraint on the origins of the
observed particles comes
from the large energy losses by hadrons through inelastic collisions 
with photons. This is also an important limit to acceleration models, 
as we shall see. Collisions between protons and photons can
generate leptonic pairs if total energy in the center of momentum
frame exceeds the proton mass by more than the pair mass.
The proton energy threshold, $E_{\pm}$ for this photopair production when
the incident photon energy is $\epsilon_{\gamma}$ is approximately
$E_{\pm} \approx 10^3/E_{\gamma}(eV)$ GeV. For the CMB $\epsilon_{\gamma}\sim 10^{-3}$eV,
so the photopair production threshold energy is about 0.01 EeV. 

The losses are much more
serious, however, if the collision energy is enough to excite the
$\Delta^+$ resonance near 1.2 GeV in the center of momentum frame, 
since that decays to produce
pions, typically carrying away of order 10\% of the total energy. 
This threshold is roughly $E_{\pi}\sim 5\times 10^7/\epsilon_{\gamma}(eV)$ GeV. 
For CMB photons the effective threshold is just below 100 EeV.
The cross section rises sharply, so that
protons near the threshold lose most of their energy in the CMB 
in less than a few $\times 10^8$ years. Thus their effective
propagation length is less than about 100 Mpc.
The significance of this was pointed out in the 1960s independently
by Greisen (1966)
and by Zatsepin \& Kuz'min (1966). If UHECRs
are of cosmological origins the result should be a relative absence of
particles above $\sim$ 100 EeV. If the source spectrum is
sufficiently flat above this ``GZK'' feature, one might also expect
a flux pile up around 100 EeV as well. The experimental
situation is unclear at the moment, since groups using complementary
techniques for analyzing extensive air showers have obtained 
conflicting spectral properties above 100 EeV.
The HiRes experiment, using atmospheric fluorescence measurements
seems to see the GZK feature (Abu-Zayyad, \etal 2002). The AGASA
muon detector reports a spectrum that actually hardens towards the
highest energies (Takeda, \etal 2002). I will consider this an open
question in the following discussions. The Auger Observatory, which
will apply both techniques simultaneously and will have
a substantially enhanced effective target area, should resolve this 
question in the next several years, in any case.

A related propagation issue is whether the UHECR events point back close to
their sources; that is, whether intergalactic magnetic fields have
any appreciable influence on UHECR trajectories. This obviously
determines the
degree to which the observed distribution mimics the source distribution.
Magnetic fields
in extragalactic space are not well known. In some rich clusters there
is evidence for fields of at least several tenths of a micro Gauss, but
elsewhere they should be much weaker (e.g., Kronberg 2001). The gyroradius can be written
conveniently as $r_g \approx E_{ZeV}/(Z B_{nG})$ Gpc, where B is now
expressed in nano Gauss. Except in clusters or possibly the vicinity of
galaxies the expected deflections should be small over distances
that hadronic CRs can propagate without severe energy losses. 
Therefore most UHECR explanations anticipate that detected
events point back to the original sources, which must then be
relatively isotropic. Some analyses have suggested enhancements in the UHECR
distribution towards ``interesting'' regions of the sky, particularly the
super galactic plane, but at present the evidence does not seem
to support any such biases (e.g., Takeda, \etal 1999; Uchihori, \etal 2000). 

The possibility that particle trajectories are
significantly deflected is still an open question,
however (e.g., Medina-Tanco, \etal 1997; Lemoine, \etal 1997;
Biermann, \etal 2001; Medina-Tanco \& En$\ss$lin 2001). Biermann, \etal 2001,
for example, have argued that a wind from our
galaxy could carry a strong enough magnetic field to sufficient
distance that UHECR trajectories would be strongly modified.
They have shown that such a field might redistribute
the 13 highest energy events detected at that time in a way
consistent with their common origin being within 
the Virgo cluster.  (See the lecture in this volume by Biermann for
more details.) On the other hand, the existence of this field structure is not
established by other means, so it remains hypothetical. 
We may hope that greatly improved UHECR statistics in the next
several years will either define unique event patterns or exclude them
from consideration.

Incidentally, strong deflections of particle trajectories
also add to the path lengths, so increase the influence of the
photopion production losses above the GZK feature unless the
sources are particularly close by. (Virgo is only about 18 Mpc away, so
this is not much of an issue in that case.) Since the path lengths
become energy dependent, it would also spread out the arrival times
of simultaneously emitted particles with the highest energy
particles arriving first.

There is at least one other basic constraint we can establish for the UHECR
sources from the limited information already established. The
measured flux combined with the energy loss rates establishes
an estimate for the energetics of the associated phenomenology.
The flux at 100 EeV translates into a local energy density of
about $3\times 10^{-22}$ J m$^{-3}$. Taking the loss time
as $3\times 10^8$ yr and assuming a steady state, the
mean luminosity of UHECRs per unit volume must be about $10^{-37}$ W m$^{-3}$
or about $3\times 10^{44}$ erg Mpc$^{-3}$ yr$^{-1}$. That is comparable
to the estimated mean volume luminosity of Gamma Ray Bursts,
a fraction of one moderate Active Galactic Nucleus (AGN) inside
100 Mpc, or the power dissipated by the accretion shock of a 
cluster like Coma or Perseus, as examples of phenomena that have been called upon to
explain UHECRs.

In summary, it seems likely that UHECRs are extragalactic in origin. If
the GZK feature is not evident then the dominant sources should be
within about 100 Mpc. If the feature is seen, and especially if
there is a pile up around 100 EeV, then a more extended distribution
of sources becomes more likely. There are
a variety of phenomena that may be energetic enough to produce
UHECRs. It is unclear yet if the distribution of
events on the sky represents the distribution of sources or has been
significantly modified by extragalactic magnetic fields. 
With these factors in mind let now look at some of the basic
physics constraints that must be satisfied by successful accelerators
of UHECRs. 

\section{How to Make a Cosmic Zevatron}                 

Accelerating a charged particle to ZeV energies is not easy! We can
express how difficult it is in a way that is almost independent of
the detailed physics of the acceleration process. I will illustrate
this point with several lines of reasoning that lead to a common
relationship first emphasized by Hillas (1984). 

The simplest constraint would be that the accelerated particles
must be contained inside the accelerator while they are being
accelerated. So, we must demand that the particle gyro radius be
less than the size of the accelerator. That is
\begin{equation}
r_g = \frac{pc}{ZeB} = \frac{E}{ZeB} < R,
\end{equation}
where $E$ is the particle energy, or
\begin{equation}
E < ZeBR = 0.9 Z B_G R_{pc} Z~~{\rm ZeV}
\end{equation}
where $R$ is a characteristic size for the accelerator. 

Magnetic fields cannot accelerate charged particles; only electric
fields can do that. It is not always easy to express the physics of
the accelerator in terms of an applied electric field, but if it
is, then obviously the maximum energy that can be achieved is given
by $E < Ze{\sf E}R$, where {\sf E} is the electric field. For
a unipolar inductor, for example, ${\sf E} \sim v_{acc} B/c = \beta_{acc} B$,
where $v_{acc}$ is a characteristic large scale speed (e.g., $\Omega \times R$).
This argument leads to the expression
\begin{equation}
E < \beta_{acc} ZeBR,
\end{equation}
which differs from the relation (2) only by the velocity factor, $\beta_{acc}$.

A related argument is that the time necessary to accelerate a particle
must be less than the lifetime of the accelerator. Let us apply this
to the diffusive shock acceleration process to be outlined in \S 4.
The mean time to accelerate a particle to energy $E$ is given by equation 9, 
so we have
\begin{equation}
t_{acc} \sim \frac{C_d\kappa}{v^2_{shock}} < t_{dynamic} \sim \frac{R}{v_{shock}},
\end{equation}
where $\kappa$ is a characteristic spatial diffusion coefficient for
the highest energy particles to be accelerated, and $C_d \sim 10$ is
a numerical factor. Under most circumstances
the applicable diffusion
coefficient is $\kappa \sim (1/3) \lambda c$, where $\lambda$ is
the scattering length of the particles. If we
require that the scattering length exceed the particle gyroradius
we once again obtain relation (3) above within a numerical factor
of order unity.

I will simply call relation (3) the ``Hillas constraint''. As did Hillas,
we can quickly filter phenomena from our discussion of viable
UHECR candidate sources with a plot of this relation. 
The line in Figure 1 shows the relationship
between source size and magnetic field that would be given by an
equality in relation (3) for a particle of energy 100 EeV/($Z\beta_{acc})$.
For comparison I have marked rough coordinates for some energetic
phenomena that are thought likely to produce cosmic rays in some form.
By the arguments given, only points above the line are viable 
candidates to produce 100 EeV CRs. This excludes supernova remnants or
winds from starburst galaxies, for example. They simply do not
have the combined size and magnetic field to produce UHECRs.

\begin{figure}
\vspace{12cm}
\includegraphics{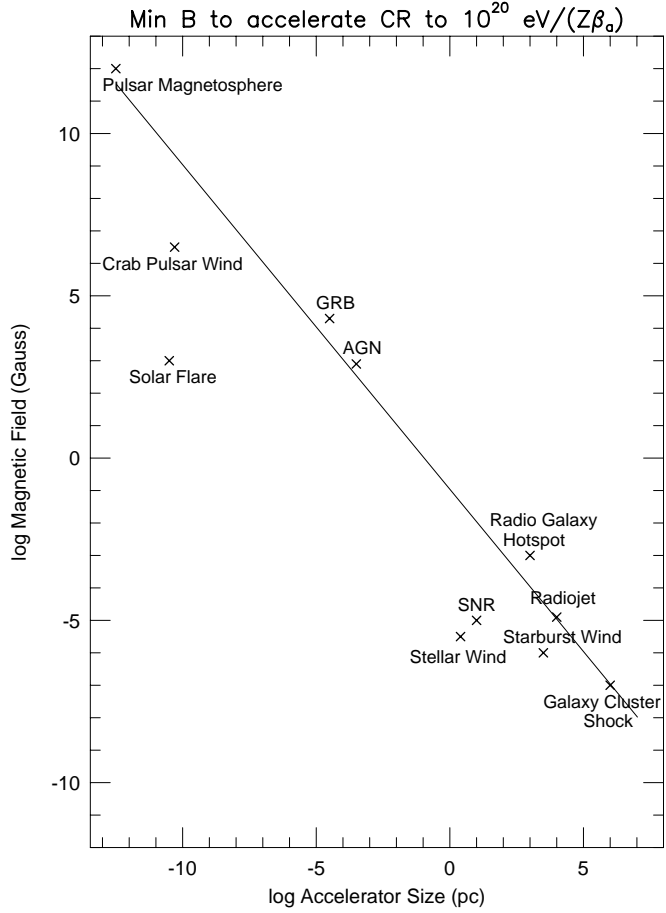}
\caption{A ``Hillas Plot'', illustrating as a function of
physical size the minimum magnetic field
required to accelerate particles of charge $Ze$ to 100 EeV. Successful
accelerators must fall above the line. Some rough coordinates for
different astrophysical objects are marked for comparison.}
\end{figure}

There are further basic filters that must be applied to phenomena
passing this first test, starting with 
limitations from energy losses within the accelerator.
In particular the energy loss times must obviously exceed
the maximum of the acceleration time and the escape time 
from the accelerator.
The two energy loss processes that most commonly limit acceleration
to the highest energies are synchrotron emission and
photon scattering. Inverse Compton scattering (elastic photon scattering)
looks much like synchrotron emission, so is usually included in that
context. Synchrotron/Compton energy losses scale as $E^2$, so
become smoothly more important as the energy increases. The energy
loss rate from inelastic collisions rises sharply in two stages
due first to photopair production and second to photopion production, 
then remains roughly constant,
with the thresholds scaling with $\epsilon_{\gamma}^{-1}$
as described in \S 2 (see also, for example, Stecker 2002, Berezinsky, \etal 2001). 

The energy loss lifetime of a particle from synchrotron/Compton
emission is simply
\begin{equation}
t_{s,c} = \frac{3}{2}\frac{m^4 c^7}{Z^4 e^4 (B^2 + B^2_{eq})}\frac{1}{E},
\end{equation}
where $m$ is the particle mass and $B_{eq} = \sqrt{6\pi u_{rad}}$ is
sometimes called the equivalent isotropic magnetic field for the photon
field in terms of Compton losses. For a black body radiation field
$B_{eq} = 0.38 T^2~\mu$G. 

Inelastic scattering losses are usually dominated by photopion
production, which turns on roughly above $E_{\pi} \sim 100~$EeV/T for
a black body radiation field. We can approximately model the energy loss
lifetime above this threshold as 
\begin{equation}
t_{\pi} \sim 3\times 10^{16}\frac{1}{T^3}~ {\rm sec}.
\end{equation}

The simplest constraint, with the escape time set to the free
flight time, R/c, is
\begin{equation}
\min(t_{\pi},t_{s,c}) > \max(t_{acc},R/c).
\end{equation}

These limitations can become effective under a variety of circumstances, but
especially when the particle accelerator is associated 
with an intense magnetic field or radiation field (i.e., ``hot'' in the black body 
sense) or if the escape time or acceleration time is long. The former
condition applies, for example, to AGNs or to young pulsar magnetospheres.
In a $10^7$ K radiation field around a young neutron star
TeV protons can propagate only a few km, so cannot escape the magnetosphere 
or be accelerated to still higher energies (Venkatesan, \etal 1997).
Similarly the radiation field near an accreting massive black hole
in an AGN will limit the escape energy of protons to less than a
few hundred TeV. In either case, accelerator regions further out
from the central massive object remain viable by this condition.

Models of Gamma Ray Bursts (GRBs) typically call for very strong magnetic
fields in equipartition with the radiation field (e.g., M\'esz\'aros \& Rees 1994). 
Then synchrotron losses are particularly important and can provide a strong limitation to
the energies of escaping particles, which may still come close enough to
ZeV to make them interesting. Adiabatic losses would also be
severe if the acceleration took place only at very early times in
the expansion of the fireball, although conditions may be suitable
long enough to avoid this issue.

At the other extreme of source environments, photopion energy losses probably eliminate
cosmic structure shocks as viable candidates for UHECRs, at least
above the GZK feature (Norman, \etal 1995; Kang, \etal 1996; Kang, \etal 1997). 
The difficulty here is the long
time required for particles to be accelerated to these energies. These
shocks have characteristic inflow speeds of only $\lsim 0.01$c, and
the magnetic fields are probably $\lsim 1~\mu$G. Those lead to
acceleration timescales of at least a few times $10^8$ years near
100 EeV, and that will be recalled as the limiting propagation
time through the CMB associated with the GZK feature.

These various constraints have significantly narrowed the field
of candidate phenomena, but there are still several interesting
possibilities. On the other hand we have so far only established
constraints and said nothing about the physics of the
acceleration process itself. In the next section I will remedy
this, focusing especially on acceleration at shocks, since
that mechanism is the one most commonly invoked to produce
high energy charged particles.

\section{Particle Acceleration Physics}

\subsection{General Comments}

Cosmic Rays are usually discussed as a distinct population of
particles in the universe, although that is largely a consequence of
the distinct ways they reveal themselves to us. This  approach
to the physics is certainly appropriate if UHECRs, for example, are byproducts of
the decay of relic massive particles or accelerated by large scale
electric fields, in pulsars, for instance. However, a somewhat more
realistic way to view these particles under most
circumstances is as a high energy component of the 
ionized matter in the universe. Galactic CRs, for example, have
roughly the same elemental composition as the interstellar medium
at those energies where this property is well measured. They are
very likely to have come from the interstellar medium.

Cosmic plasmas are largely ``collisionless''
in the sense that binary Coulomb collisions play a minor role in
dissipation of energy. Instead, dissipation comes from ``collective'' 
interactions that depend on fluctuations in charge and currents (i.e., plasma
waves or turbulence). These interactions tend to be selective in their
efficiency, so that cosmic plasmas are probably rarely fully
thermalized. The problem is that they never have the time and they are not
really closed systems in the thermodynamic sense. This especially 
applies to the most energetic charged particles, a point that was
also central to our discussions in the previous section. In that sense
cosmic rays can be seen as those particles that have so-far escaped being
thermalized, but have preferentially extracted energy from a
system out of thermodynamic equilibrium. The prototype acceleration
model for this was that devised by Fermi (1949). There are
several modern variants of his idea; they all involve stochastic
processes in which charged particles with effective mean free paths
too long to allow them to become thermalized extract free energy
via interaction with fluctuations in a dynamic system. 
The currently most commonly discussed
acceleration process, diffusive shock acceleration, is of this
type, and I will devote some space to its details momentarily.
I note that not all collisionless plasma models depend on Fermi
processes, however. In some other settings, such as magnetic
reconnection, acceleration is often predicted as a result of
locally strong induced electric fields. However, The microphysics of this once
again involves the details of a breakdown in thermodynamic or
bulk magnetohydrodynamic behaviors (e.g., Lesch \& Birk 1998).

\subsection{Diffusive Particle Acceleration at Shocks}

Starting from the premise above, we can understand so-called
diffusive shock acceleration (``DSA'') as a part of the
physics of collisionless plasma shocks. Cosmic shocks would
mostly not exist if they had to depend on binary interactions;
they form, instead due to very complex, nonlinear phenomena 
involving plasma turbulence (e.g., Quest 1988). 
The end result is a transition that in some respects can be
described by classical fluid shocks, but in others may not
be. Like classical shocks a region develops where most of the particles entering
from upstream interact strongly and have much of their directed
energy ``dissipated''; that is they are ``thermalized''.
In this context we will identify CRs as those particles that
are not thermalized because they interact too weakly with the
plasma turbulence to become trapped inside the primary dissipative
layer. 
The dissipative layer is usually called the shock, 
although it is only part of the shock
structure, so I will use the more appropriate ``subshock'' label. 
The thickness of the subshock is not easily characterized,
and it depends on the orientation of the local mean magnetic field.
Often it is described in terms of the characteristic gyroradius of 
thermalized particles.

If we can ignore the mass, energy and momentum carried
by the CRs then the usual hydrodynamic or MHD shock jump conditions apply
across the subshock. CRs then can be treated as test particles
interacting with this flow. The central idea behind DSA is that
these CRs still are sufficiently strongly scattered by local plasma 
turbulence that their motions become randomized (but not thermalized)
close to the subshock. The resulting behavior is fairly easily 
described if the speed of the CRs is large compared to the bulk motions
relative to the shock and if we can approximate the
distribution function, $f(p)$, as almost isotropic in
both the upstream and downstream frames of reference. In that case, for example, the flux of
CRs crossing the subshock from upstream in a momentum range dp
is approximately $\pi v p^2dp f(p)$,where $v$ is
the particle speed. On the other hand, 
the CRs downstream are advected away at the
bulk flow speed of the postshock plasma, which we can call $u_2$.
The flux there is $4\pi u_2 p^2dp f(p)$.
In a steady state the ratio of these two fluxes determines the
probability that a particle entering the shock from upstream
will escape downstream at the same momentum, ${\sf P}_e$. Since
the steady downstream distribution function must be
uniform and also cannot change across the subshock (the
CRs ignore the subshock, remember), this gives ${\sf P}_e \approx 4u_2/v << 1$.
Thus, under these assumptions most of the CRs are temporarily trapped between
the upstream and the downstream flows. Also, by construction they
are not stopped within the subshock itself, so that they are akin to 
ping pong balls trapped between converging paddles, here
represented by the large scale plasma turbulence. It is simple
to show that on average each time they cycle from upstream to downstream
and back again they gain momentum given by 
\begin{equation}
\frac{\delta p}{p} \approx \frac{4}{3} \frac{u_1 - u_2}{v} = \alpha <<1,
\end{equation}
where $u_1$ represents the upstream bulk plasma speed into the shock.
Once the particles are relativistic the fractional momentum
gain, $\alpha$ is a constant, as is the escape probability, ${\sf P}_e$.

Under such circumstances Fermi showed that the steady state momentum
distribution is a powerlaw (since there is no associated momentum
scale) given by $f(p) \propto p^{-q}$, with $q = 3 + \frac{{\sf P}_e}{\alpha}$.
Fermi was not thinking of shocks, but in the late 1970s several people 
independently recognized how this Fermi-like process would work in 
shocks (Axford, \etal 1977; Krymskii 1977; Bell 1978; Blandford \& Ostriker 1978).
In that case the powerlaw index of the momentum distribution becomes
$q = \frac{3u_1}{u_1 - u_2} = \frac{3r}{r-1},$ where $r = u_1/u_2 = \rho_2/\rho_1$,
is the compression through the shock.
For a strong, nonrelativistic adiabatic shock $r\rightarrow 4$,
so $q\rightarrow 4^+$. For even moderately strong shocks the
powerlaw slope is close to 4, so the solution appears pretty robust.
Notice that this result depends only on the jump conditions across
the shock, and it was necessary only to suppose that somehow
plasma turbulence would isotropize the CRs on each side of the shock.
Plasma shocks were long known to be highly turbulent downstream. 
The other key insight to the introduction of DSA was that unless
the CRs returning to the upstream side of a shock where isotropized
with respect to the upstream flow they would amplify local
Alf\'venic turbulence there by a streaming instability. Those waves
should then, in turn scatter the subsequent CRs trying to escape in
this way. This physics, even though it leads to a simple outcome, actually
shows that DSA is a nonlinear aspect of shock formation, so
was an early clue that the physics is actually more complex.

An important detail about DSA that we encountered
in \S 3 was the finite amount of time required to
accelerate particles up to a specified energy (or momentum). This 
is now easy to estimate from the fractional momentum
gain per shock encounter expressed in equation 8 along
with an estimate of the time interval between CR shock crossings.
The CRs propagate approximately a diffusion  length, $x_d = \kappa/u$,
on each side of the shock before their return. Since the
CRs move at a speed $v$, the average time between a pair
of shock crossings is $2(x_d(1) + x_d(2))/v$, where the
subscripts refer to the upstream and downstream values for
$x_d$. From this the mean acceleration time is (e.g., Lagage \& Cesarsky 1983)
\begin{equation}
t_{acc} = \frac{3}{u_1-u_2}\int^{p_{max}}_{p_{min}} 
\left [ \frac{\kappa_1}{u_1} + \frac{\kappa_2}{u_2}\right ] \frac{dp}{p}.
\end{equation}
This expression has been verified by a variety of numerical
approaches. In particular it provides a reasonable
estimate of the upper cutoff to the CR spectrum one expects
if the particles are injected locally at the shock.

The most obvious nonlinear complication in DSA physics comes from
a calculation of the pressure associated with the CRs. That is
\begin{equation}
P_{cr} = \frac{4}{3} \int^{p_{max}}_{p_{min}} v p^4 f(p) d\ln p.
\end{equation}
It is clear that for the strong shock limit ($q = 4$) $P_{cr}$
diverges logarithmically with $p_{max}$. Now,
that momentum will in practice always be limited by some
constraint, such as those we have discussed. Still, it
is apparent that the pressure in CRs can potentially compete
with the thermalized plasma or even dominate it. This is quite an important
realization, especially since the CRs diffuse upstream from
the subshock a characteristic length $x_d$ as part of the acceleration process. 
That means there will be a pressure
gradient $\sim P_{cr}/x_d$ influencing the flow upstream,
so that the flow into the subshock will in this
``shock precursor'' be fractionally reduced
by an amount $\delta u/u_1 \sim P_{cr}/(\rho_1 u^2_1)$.
The maximum postshock thermalized pressure is 
$\frac{3}{4}\rho_1 u^2_1$, so if the CR and thermal pressures
expected at the subshock are comparable, the shock
must be significantly modified from the classical
hydrodynamic shock. 

The CR feedback generally tends to enhance
the dynamical role of the CRs, since the deceleration through 
the precursor weakens the subshock and thus reduces the
postshock thermal pressure. At the same time it turns out
that the total compression through the shock, including
the precursor and subshock generally exceeds that for a
purely hydrodynamical shock. Then the expected value of $q$
is reduced for higher energy CRs, which tend to have longer scattering
lengths, so see a larger compression. 
That adds to the CR pressure as well, since the
relative number of the most energetic particles is greater. At the same
time, if the highest energy
CRs escape upstream into a region without sufficient
scattering to return them they remove energy (and total pressure)
from the shock. This likely eventuality further enhances the total
compression through the shock, much like what happens in a
radiative shock. On the other hand, some models explaining
how particles manage to be ``injected'' into the CR population
by escaping upstream through the shock for the first time
would predict a reduced injection efficiency once the subshock
is weakened. That, in turn, would reduce the expected $P_{cr}$,
so provide feedback of the opposite sign. Malkov, \etal (2000) have shown
a bifurcation in the steady state analytic solutions when these
various features are included and argue that a critical, self-organization
may develop that determines the weights of the various nonlinear
effects.
Figure 2, from Kang, \etal 2002, illustrates some these
features in the evolution of a simulated CR shock.

\begin{figure}
\vspace{18cm}
\includegraphics{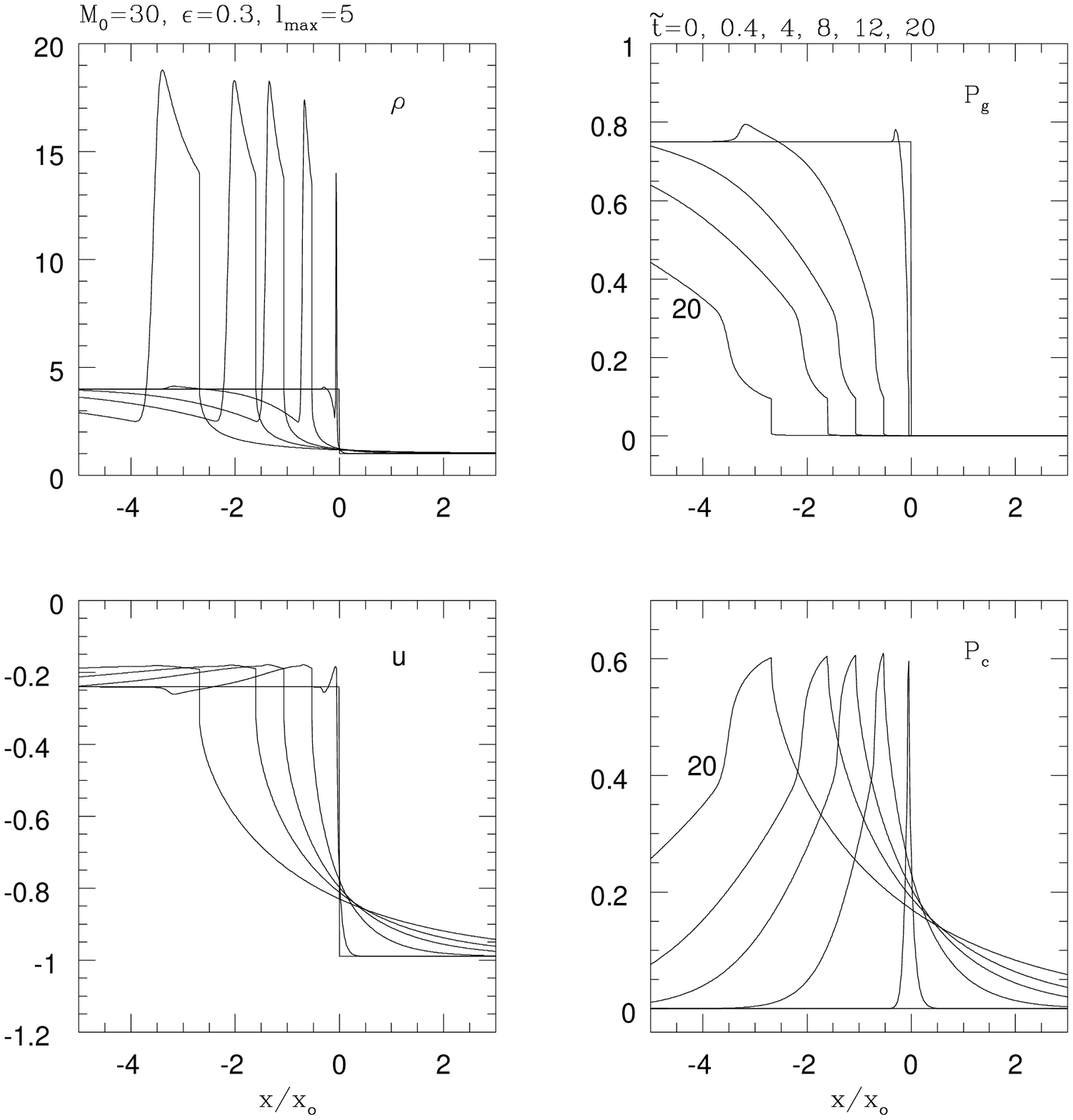}
\includegraphics{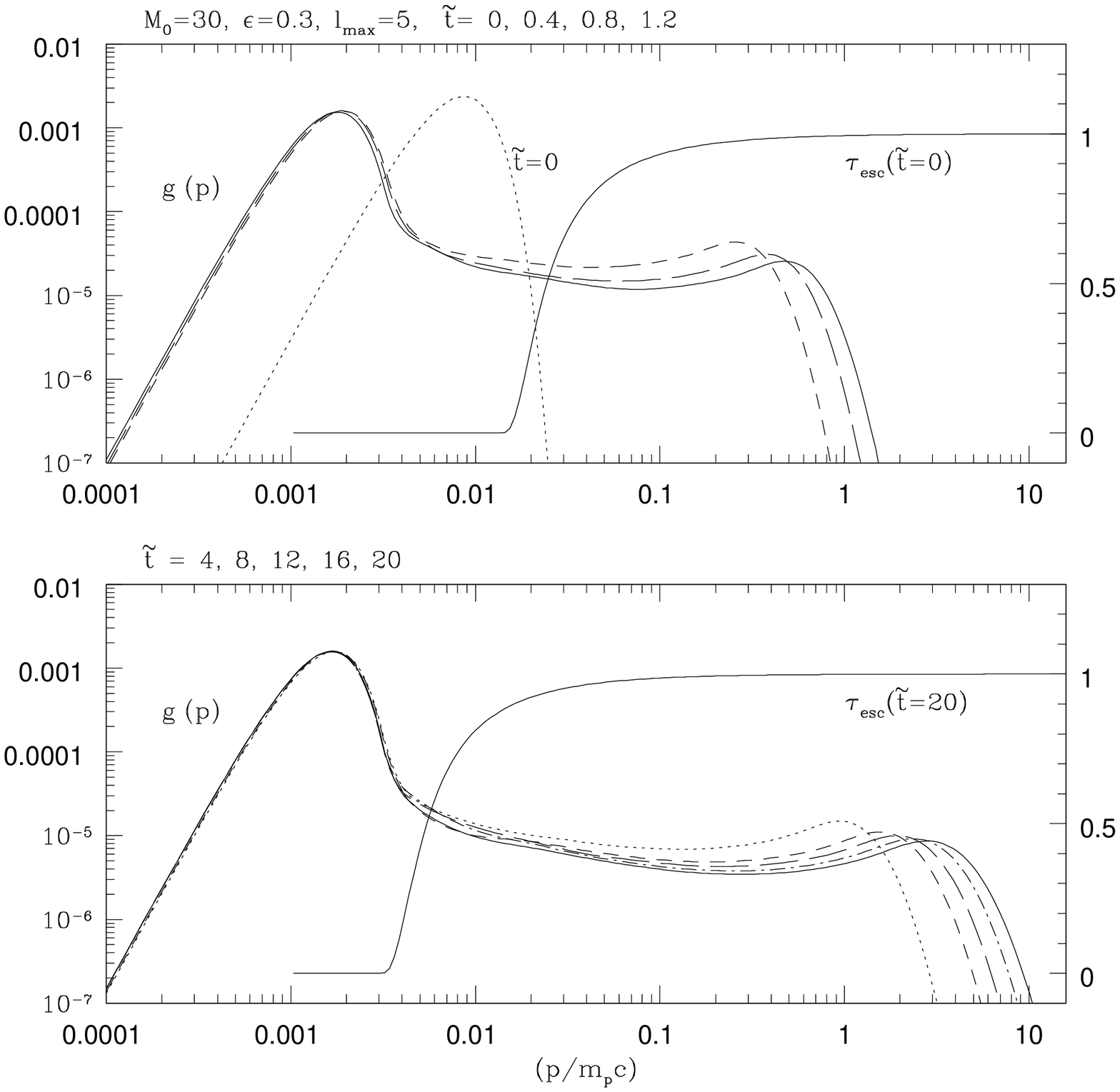}
\caption{Evolution of a simulated CR modified shock with initial
Mach number 30. The upper four panels show evolution of dynamical
variables, including gas density, pressure and flow velocity, along with 
the CR pressure. The initial conditions were a simple gasdynamic shock,
visible as a step function in the plots. The bottom two panels show
the evolution of the proton momentum distribution, shown here as
$g(p) = p^4 f(p)$. The spatial diffusion coefficient was Bohm
diffusion and the times are given in units of the diffusion time
for p = mc.}
\end{figure}

There are still other nonlinear
features of the DSA process, most of which are
less well understood than what I just described.
I refer interested readers to some of the recent
reviews for further information (e.g., Malkov \& Drury 2001;
Jones 2001).

Especially in some phenomena considered for acceleration
of UHECRs the flows into the shocks may be
relativistic or even ultrarelativistic. This changes
important details about DSA that make
its results harder to calculate. First, the
assumption that we could consider the distribution
function, $f(p)$, to be isotropic in both the
upstream and downstream frames obviously cannot
apply, since the bulk speeds are no longer 
a small fraction of the CR speed. Second, an
unstated detail in my previous description 
was that the orientation of the large scale
magnetic field relative to the shock normal
could be ignored to a first approximation.
However, especially for relativistic shocks
an oblique field may intersect the
shock face ``superluminally''; that is, the
intersection moves along the face at
superluminal speed. Particles cannot then re cross
the shock from downstream by following an individual field 
line. In that case only
motions of particles across field lines are
relevant to DSA and that is more difficult
to model. Despite these difficulties, the test particle form of the
process is reasonably well understood now,
and is nicely discussed in recent literature
(e.g., Achterberg, \etal 2001). Here I mention only a
couple of important comparisons with
nonrelativistic DSA properties. On the
face of it one might expect DSA to be somehow more
efficient in a relativistic shock, since the
fractional energy gain per crossing is large, $\propto \Gamma^2$. 
where $\Gamma$ is the Lorentz factor of the shock. On the
other hand, the compression ratio of a highly relativistic
shock is less than that for a nonrelativistic shock, 
asymptoting to 3 rather than 4. This enhances the
relative escape probability of the CRs, so
leads to a steeper momentum distribution. For asymptotically strong
relativistic shocks the expected value of $q \approx 4.2$ rather than
$q \approx 4$. Thus, compared to nonrelativistic shocks we expect
reduced nonlinear feedback of from CRs (still defined
as those particles that are not thermalized).

Before returning to discuss astrophysical models for UHECR accelerators
it is helpful to mention another proposed acceleration mechanism that
is similar to DSA, but which would operate across thin shear layers
rather than across shocks. It was pointed out by Jokipii, \etal (1989)
that diffusion of CRs across a shear layer would produce similar
results to a shock provided the CR mean free paths exceed the thickness
of the shear layer. Once again the CRs find themselves caught in an apparently
convergent flow. In nonrelativistic flows the process is relatively
inefficient, but Ostrowksi (2000) has argued that the large boosts
that would accompany this process in the boundary of a relativistic jet
might make it astrophysically interesting. 

\section{Comments on Some Candidate Zevatron Models}

Our discussions in \S 3 considerably limited the range of candidate
UHECR accelerators.
While there are currently a variety of UHECR accelerator scenarios
on the market with serious supporters, I will focus here on only
three that are illustrative. The most viable candidate in my view
may be high powered jets from AGNs interacting with the
intergalactic medium (IGM). Gamma Ray Bursts may be excluded on the basis
of their redshift distributions, as I will mention, yet 
they may be physically just capable of producing such particles, so I
will include them. Since the possibility that UHECRs are heavy ions
is not yet excluded, I will also include one example of such models;
namely, acceleration of Fe ions in young pulsar winds.

\subsection{Giant Radio Jets}

AGNs are thought to derive their tremendous energy supplies from accretion onto
supermassive black holes. We excluded the AGN environment itself from
the UHECR discussion on account of the severe losses expected from
photopion production there. However, a significant subpopulation of AGNs
expel large amounts of energy in directed outflows, typically
identified through associated radio emissions. These so-called
radio jets are probably formed as relativistic beams near the
black hole and then drill their way to large distances. In some cases
the interaction regions extend as much as a Mpc from the galaxy itself.
The energy of the jet is dissipated in a bow shock inside the IGM
and in shocks within the jet plasma itself. 
The bow shock is probably not fast enough to produce UHECRs.
In the cartoon model
of radio jets there is a strong termination shock at the end of the jet.
The characteristic size of such a shock would be $R \sim 10$kpc;
the magnetic field (from radio and X-ray observations) might be
$B \sim 10 - 100 \mu$G, and $v_{shock} > 0.1$c.
With these numbers the Hillas constraint for protons would be $E_{max} \sim 1$ ZeV
(e.g., Rachen \& Biermann 1993).
The acceleration timescale $\sim 10^7$ yrs, which is comfortably
within the expected lifetime of the jets. The constraints
from synchrotron and photopion losses give similar limits.
Simulations show that the cartoon picture of radio jets
is too simple (e.g., Tregillis \etal 2001), since the
termination shock is unsteady and much of the dissipation actually
involves a sequence of somewhat weaker shocks. Nonetheless, the
shocks in the jet are large and very fast, so the radio jet
scenario remains viable in my view. In addition, Ostrowski has
argued that if the relativistic jet has a thin boundary layer
particles may be accelerated quickly by the diffusive shear
acceleration process mentioned in \S 4. I am not convinced that
such thin layers will exist, but perhaps this could contribute. Other
authors have suggested that magnetic reconnection associated with
the jet flows could also play a role (Schopper, \etal 2001). 

One important detail here concerns the composition of radio jets.
It is unclear if there is a substantial proton component in these
jets. Energy requirements are lower if the jets are pair
plasmas and there are observational indications that the composition nearer
the AGN is dominated by pairs (Wardle, \etal 1998). On the other hand it is
difficult to create the jets entirely with pairs, because of
extreme energy losses, and some models posit that the baryon
load is dominant (e.g., Mannheim 1993; Falcke \& Biermann 1995).

While they are very powerful and may be capable of accelerating UHECRs,
radio jets are quite rare phenomena, especially in the local universe.
There are a few candidates inside the 100 Mpc range associated with
the GZK feature. Centaurus A is only about 3 Mpc away, for example, and
M87 is about 18 Mpc from us. Both of these are relatively low
luminosity objects, but might be adequate. Really high luminosity
radio jets, such as Cygnus A ($\sim 200$ Mpc) or quasars are
generally much farther away. The difficulty with sparse sources,
as mentioned in \S 2, is how to produce an apparently isotropic distribution
of detected events. Early indications for concentrations along
the supergalactic plane, which might signal association with
matter concentrations, seem to have gone away with increased data (Takeda,
\etal 1999; Uchihori, \etal 2000). If the
GZK feature is seen, this may open the door to a larger scale
distribution. The alternative possibility may still be open
that intergalactic magnetic fields able to redistribute the events.

\subsection{Gamma Ray Bursts}

Gamma Ray Bursts are an attractive idea for UHECRs, because they
involve highly relativistic flows and possibly very strong shocks.
The more common long-duration burst events are thought now probably to represent
the formation of a black hole during core collapse of a massive 
star (M\'esz\'aros 2002), with evidence pointing towards the generation of
relativistic jets, perhaps with Lorentz factors of several hundred.
These high Lorentz factors can result when very little mass is
incorporated ($<10^{-5}~ {\rm M}_{sun}$), and when
the total kinetic energy in the flows is comparable to that in a conventional
supernova, ($\sim 10^{51}$erg).
The ejected plasmas (``fireballs'') are thought to be pair dominated, because they are
so incredibly hot, although there will be a minor baryonic
``load.''
In this scenario variations in the energy ejection are supposed to
lead to internal, mildly relativistic, shocks within the fireball, and
those are plausibly capable of accelerating protons to UHECR energies
(e.g., Waxman 2000).
We looked at the energy constraints on CRs being accelerated
in \S 3 and found that energies near the UHECR range were barely
possible, with the most severe constraint being set by synchrotron
losses in the posited magnetic fields.
As noted in \S 2 if they were distributed uniformly in the universe,
GRBs might be common enough to account for the measured flux of
events. However, as pointed out by Stecker (2000)
GRBs seem to be relatively rare today; they seem instead to have
been much more common in the past. In that case, their local
space density is much to small, according to Stecker, to
account for UHECRs.

\subsection{Pulsar Winds}

I argued before that UHECRs are probably extragalactic, and this is
true. On the other hand, if they could be made of iron rather than protons,
then it is perhaps still plausible to discuss galactic scenarios. The
obvious immediate worry is how to explain the absence of an associated asymmetry
in the distribution of detected events.

While protons
near 1 ZeV are not significantly deflected in their motions by the galactic
magnetic field, iron nuclei could still be influenced. At energies
below 100 EeV the distribution of events might appear roughly
isotropic, even if the sources were near the plane. 
In that spirit I will briefly outline a recent model
for UHECRs based on young pulsars.

The magnetic fields typically associated with pulsars ($B\gsim 10^{12}$G)
in concert with their rapid rotation ($\omega \lsim 10^3 {\rm s}^{-1}$)
could in principle accelerate protons to energies higher than 10 EeV,
and a factor Z = 26 higher for Fe, according to the Hillas constraint.
Magnetar fields are stronger, although the rotation periods are slower,
so the nominal limits are similar.
So, in that regard neutron stars are natural candidates for UHECRs. Venkatesan, \etal (1997)
pointed out, however, that pair production should limit the maximum
energy in the magnetosphere to about 1000 TeV, and that synchrotron
losses near the neutron star are also severe. Blasi, \etal (2000)
have argued, on the other hand, that relativistic pulsar winds
outside the light circle are capable of accelerating Fe to energies
above 100 EeV. This is allowed by the Hillas constraint and also
consistent with the energy per particle
that would be carried away in a wind extracting the rotational
energy of a rapidly rotating pulsar. Being outside the
intense photon field and magnetic fields of the near magnetosphere,
these particles should be able to escape. These authors also estimate
that the combined efficiency of acceleration and galactic trapping
need be only $\sim 10^{-5}$ to account for the observed flux,
given a birthrate of pulsars in the galaxy around $10{^-2}{\rm yr}^{-1}$.

Especially at the highest energies it is still difficult to produce
a truly isotropic distribution of detected events in such a scenario.
So, once again, the improved statistic expected in the near future
from Auger should determine if any such class of models is really
viable. 

\section{Conclusions}

I have tried to outline in a simple fashion some of the underlying issues
associated with the physics of accelerating charged particles
to energies close to 1 ZeV. The evidence is strong that
nature somehow produces such particles, either by acceleration or
by decay from something even more energetic. They are rare, but
their existence demonstrates something new. Both possible scenarios stretch
our current understandings. Here I have taken the more conventional
road supposing that physical environments exist where particles
can reach such energies from below. There are few accelerators
we have been able to imagine that apply sufficient energy gains
and contain particles well enough for them to reach these energies.
Even if that seems possible in principle, the energies losses
faced by such particles are huge, due especially to collisions with
ambient photons or by synchrotron radiation. 

The bottom line is that very few candidate environments exist. Mostly
they are extragalactic, both because it is difficult to 
confine the UHECRs in the galaxy, and because there are few galactic
candidate phenomena. New
proposals appear routinely, and some other suggestions may turn out
to be obvious. At the moment, however, it seems to me that the
strongest candidate accelerators are the high powered jets
associated with radio galaxies, although they are not sure bets. 
These jets start out as moderately
relativistic, they can carry their energy to large distances and
they have lifespans of $10^8$ years or so. There are several
plausible dissipation mechanisms that have been identified in
radio jets that may be capable of producing very high energy
particles, including shock acceleration, shear acceleration and
even magnetic reconnection. We know from observations of nonthermal
X-rays in some such objects that electrons have been 
generated with energies in the TeV range. UHECRs are probably
hadronic, not leptonic, and that may be an issue. There are
arguments favoring hadronic jets in these objects, although
the more common view is that they are leptonic.

No matter where this topic goes it is certain to be interesting in the
coming few years. Experiments under construction, such as the
Auger Observatory, and those under discussion for the
future, such as OWL, will vastly increase detection capabilities.
Then we should finally know with confidence what role is
played by photopion losses in UHECR propagation and how
the detected events are distributed on the sky. That
information should tell us much more clearly how the
sources must be distributed, what their energy spectra
must be and perhaps, if we are lucky, something about the
propagation of the CRs, and thus information about 
intergalactic magnetic fields. Along the way we should
learn finally if UHECRs are probes of fundamental physics
or astrophysics.

\acknowledgements                                        

I am most grateful to Professor Norma S\'anchez for her invitation to
participate in the delightful and stimulating
9$^{th}$ Chalonge School on Astrofundamental Physics. My work on
the physics of particle acceleration is 
supported by the NSF through grant AST00-71167, by NASA through
grant NAG5-10774 and by the University of Minnesota Supercomputing
Institute.

\end{document}